\documentclass{WileyMSP-template}
\usepackage[english]{babel}
\usepackage{float}
\usepackage{multirow}
\usepackage{xspace}
\usepackage{graphicx}% Include figure files
\usepackage[normalem]{ulem} %allows one to crossout
\usepackage{soul} %allows one to crossout
\graphicspath{{Figure//}} %allows to omit the path in the \incudegraphics command
\usepackage{dcolumn}% Align table columns on decimal point
\usepackage{bm}% bold math
\usepackage{color}
\usepackage{hyperref}
\usepackage{array}
\usepackage{booktabs}
\usepackage{multirow}
\hypersetup{
	unicode=false,          % non-Latin characters in Acrobat’s bookmarks
	pdftoolbar=true,        % show Acrobat’s toolbar?
	pdfmenubar=true,        % show Acrobat’s menu?
	pdffitwindow=false,     % window fit to page when opened
	pdfstartview={FitH},    % fits the width of the page to the window
	pdftitle={My title},    % title
	pdfauthor={Author},     % author
	pdfsubject={Subject},   % subject of the document
	pdfcreator={Creator},   % creator of the document
	pdfproducer={Producer}, % producer of the document
	pdfkeywords={keyword1} {key2} {key3}, % list of keywords
	pdfnewwindow=true,      % links in new window
	colorlinks=true,       % false: boxed links; true: colored links
	linkcolor=blue,          % color of internal links (change box color with linkbordercolor)
	citecolor=blue,        % color of links to bibliography
	filecolor=blue,      % color of file links
	urlcolor=blue,       % color of external links
	plainpages=false        % influences the page numbering
}

\newcommand{\red}[1]{\textcolor{red}{#1}}

\newcommand{\Tc}{$T_{\rm c}$\xspace}%superconducting Tc
\newcommand{\TC}{$T_{\rm C}$\xspace}%Curie temperature

\newcommand{\Alg}{$A_{1g}$}

\newcommand{\AZg}{$\rm{A_{2g}}$\xspace}

\newcommand{\Eg}{$E_{g}$}

\newcommand{\FS}{\mbox{Fe$_3$Sn$_2$}\xspace}
\newcommand{\wn}{\mbox{$\,\mathrm{cm^{-1}}$ }}               % wavenumber

\newcommand{\vk}{\textbf{k}}
\newcommand{\vq}{\textbf{q}}

\let\red\relax

\begin{document}

\pagestyle{fancy}
\rhead{\includegraphics[width=2.5cm]{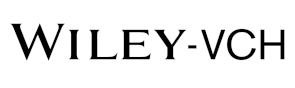}}

\title{Phonon anomalies associated with spin reorientation in the\\
Kagome ferromagnet \FS}

\maketitle

% Author: Please give full first and last names for authors and include * after the name of all corresponding authors

\author{Ge He}
\author{Leander Peis}
\author{Ramona Stumberger}
\author{Lilian Prodan}
\author{Vladimir Tsurkan}
\author{Nico Unglert}\\
\author{Liviu Chioncel}
\author{Istv\'an K\'ezsm\'arki}
\author{Rudi Hackl}

% Dedication

\dedication{}%Optional dedication here. If no dedication is required, please leave blank}

% Affiliations: Please provide adacemic titles (Prof. or Dr.) for all authors where applicable, and include an institutional email address for all corresponding authors
\begin{affiliations}
Dr. Ge He, Leander Peis\\
Walther Meissner Institut, Bayerische Akademie der Wissenschaften, 85748 Garching, Germany\\
Nico Unglert, Prof. Dr. Liviu Chioncel\\
Theoretische Physik III, Institut f\"ur Physik, Universit\"at Augsburg, Universitätsstrasse 1 (Physik Süd), 86159 Augsburg, Germany\\
Dr. Lilian Prodan, Dr. Vladimir Tsurkan, Prof. Dr. Istv\'an K\'ezsm\'arki\\
Experimentalphysik V, Institut f\"ur Physik, Universit\"at Augsburg, Universitätsstrasse 1 (Physik Süd), 86159 Augsburg, Germany\\
Ramona Stumberger, Prof. Dr. Rudi Hackl\\
Technische Universit\"at M\"unchen, Fakult\"at f\"ur Physik, 85748 Garching, Germany\\
Walther Meissner Institut, Bayerische Akademie der Wissenschaften, 85748 Garching, Germany\\
Email Address: hackl@tum.de\\

\end{affiliations}

% Keywords: Please provide a minimum of three and a maximum of seven keywords, separated by commas

\keywords{Phonons, light scattering, magnetism}

% Abstract should be written in the present tense and impersonal style (i.e., avoid we), and be at most 200 words long
\begin{abstract}

Polarization- and temperature-dependent Raman data along with theoretical simulations are presented for the Kagome ferromagnet \FS. Eight out of nine expected phonon modes were identified. The experimental energies compare well with those from the simulations. The analysis of the line widths indicates relatively strong phonon-phonon coupling in the range 0.1 to 1. The temperature-dependent frequencies of three A$_{1g}$ modes show weak anomalies at approximately 100\,K. In contrast, the linewidths of all phonon modes follow the conventional exponential broadening up to room temperature except for the softest A$_{1g}$ mode, whose width exhibits a kink close to 100\,K and becomes nearly constant for $T>100$\,K. These features are indicative of a spin reorientation taking place in the temperature range above 100\,K  \red{which might arise from spin-phonon coupling}. The low-energy part of the electronic continuum in \Eg symmetry depends strongly on temperature. The possible reasons include particle-hole excitation tracking the resistivity, a spin-dependent gap or spin fluctuations.

\end{abstract}

% Text: Please use section headings and subheadings as specified below. For communications, all section headings apart from Experimental Section should be removed
% Please make the first reference to a display item bold: \textbf{Figure 1}
% Do not abbreviate Figure, Equation, etc.; display items are always singular, i.e., Figure 1 and 2.
% Equations are always singular, i.e., Equation 1 and 2, and should be inserted using the {equation} environment, not as graphics
% Please do not use footnotes in the text, additional information can be added to the Reference list.

\section{Introduction}
\begin{figure*}[ht!]
	\centering
	\includegraphics[width=0.99\linewidth]{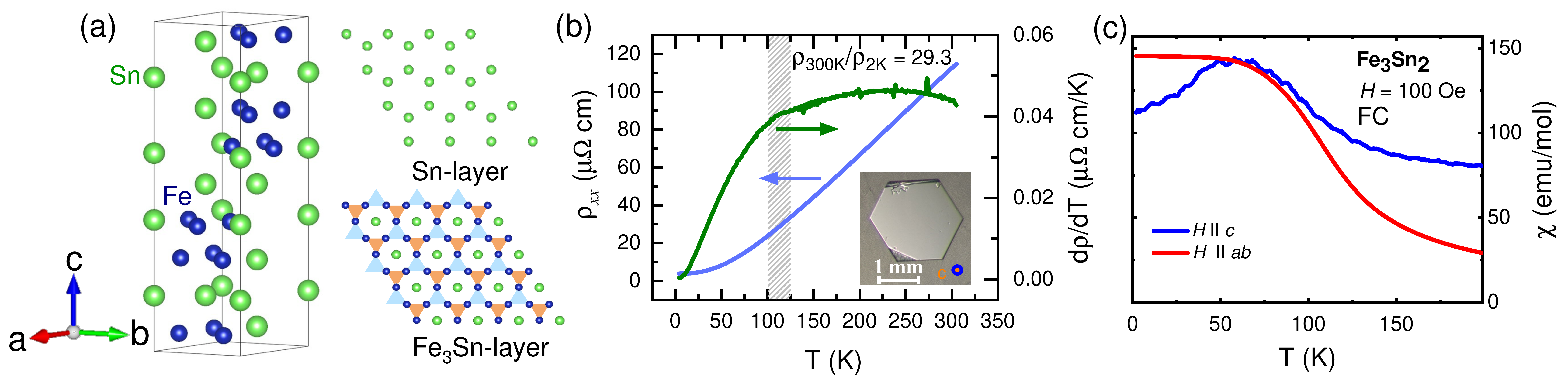}
	\caption{Properties of \FS. (a) Crystal structure (left). The crystals consist of hexagonal Sn-layers sandwiched between  Kagome Fe$_3$Sn double layers (right). (b)  Resistivity $\rho(T)$ of \FS. $\rho(T)$ (blue) was measured on a crystal from the same batch as the sample used here.  $d\rho(T)/dT$ (green, right axis) is nearly constant above 100\,K indicating a linear temperature dependence of $\rho(T)$. Below a rather well-defined kink at 100\,K the slope changes rapidly. (c) Anisotropic susceptibility $\chi_i(T)$ for field orientations as indicated. The sample was cooled in an applied field of $\mu_0 H=0.01$\,T. $\chi_{ab}(T)$ increases monotonously and almost saturates below 50\,K. $\chi_{c}(T)$ has a maximum between 50 and 100\,K.}
	\label{fig1}
\end{figure*}

Materials with novel properties and functionalities have intriguing scientific and applied perspectives. In particular magnets with exotic ground states continue to attract attention since the properties may be tailored \textit{via} the lattice and/or the electronic band structure. \FS is a layered Kagome ferromagnet with a very high Curie temperature \TC$ = 670$\,K. The Fe-Sn bi-layers are separated by Sn mono-layers \cite{Trumpy:1970,LeCaer:1978,Malaman:1978,fenner_non-collinearity_2009} as shown in Fig.~\ref{fig1}(a). Due to its out-of-plane anisotropy \cite{LeCaer:1978,Malaman:1978} with the magnetic easy axis perpendicular to the Kagome layers, magnetic stripes and a variety of magnetic bubbles have been observed in thin lamellae using Lorentz transmission electron microscopy \cite{HouZP:2017,Hou:2018,Tang:2020a,Tang:2020b}. Since the material possesses inversion symmetry, these mesoscale magnetic textures do not form due to the antisymmetric part of the exchange coupling tensor, i.e. the Dzyaloshinskii-Moriya interaction, but due to a competition between the out-of-plane anisotropy and the magnetic dipole-dipole interaction. The magnetic anisotropy of the system is also linked to the reconstruction of its band structure in a magnetic field: A strong shift of bands close to the Fermi energy was found to be dependent not only on the magnitude but also on the orientation of the magnetic field \cite{YinJX:2018}. Moreover, indications of a temperature-driven spin reorientation, from out-of-plane at high temperatures towards in-plane at low temperatures, have been found earlier in M\"ossbauer \cite{LeCaer:1978,Malaman:1978}, x-ray \cite{fenner_non-collinearity_2009} and transport studies \cite{WangQ:2016}. Optical absorption experiments \cite{Biswas:2020} reveal an additional feature in the 10\,meV range below 150\,K and associate it with this spin reorientation. Similarly, transport and magnetisation experiments [performed on our own samples and plotted in Fig.~\ref{fig1}(b) and (c)] show a cross-over temperature close to 100\,K and will be discussed in detail in section \ref{sec:samples}.

There are various other effects in a planar hexagonal lattice such as massless Dirac fermions. As opposed to graphene the tips of the cones typically intersect below the Fermi energy $E_{\rm F}$ and may become massive in the presence of spin-orbit coupling (SOC) \cite{YinJX:2018,Lin:2020}. Recently, the existence of flat bands at approximately 200\,meV below $E_{\rm F}$ in confined regions of the Brillouin zone was associated with magnetic ordering in \FS \cite{Lin:2018}. These observations and suggestions are among the main motivations for the present work, focusing on spin reorientation and band reconstruction phenomena in \FS and its consequences for the lattice dynamics and charge response, both probed by Raman spectroscopy.

Spin textures usually entail a huge anomalous Hall effect (AHE) being associated with the Berry phase the electrons pick up upon moving across a magnetic background \cite{Nagaosa:2010}. Thus \FS has similarities with, e.g., MnSi \cite{Franz:2014d} and various other compounds. The origin, however, of the rather complex itinerant ferromagnetism with a high \TC and a spin reorientation at some 100\,K is elusive. The most popular approaches are based on the Hubbard model and either favor flat-band ferromagnetism \cite{Mielke:1991b,Mielke:1991a,Mielke:1992} or a trade-off between potential and kinetic energy \cite{Lin:2018,Nagaoka:1966,Pollmann:2008}. The latter case is reminiscent of the magnetism in Fe(Se,Te) where itinerant and nearly localized spins seem to cooperate as well \cite{Yin:2011,Leonov:2015,Stadler:2015}.

It is unlikely that the magnetism in \FS can be observed directly in a similar fashion as in FeSe since the two-magnon excitations typical for antiferromagnets \cite{Baum:2019}  do not exist here. However, indirect signatures of spin order, the spin reorientation or the interaction between spin, lattice and electrons may be expected, in particular gaps between flatbands such as in the optical experiments \cite{Biswas:2020} or phonon renormalization effects as in MnSi \cite{Eiter:2014a}. In this, to our knowledge first, Raman study of the topological material \FS we start with analyzing the phonons.

%%%%%%%%%%%%%%%%%%%%%%%%%%%%%%%%%%%%%%%%%%%%%%%%%%%%%%%%%%%%%%%%%%%%%%%%%%%%%%%%%%%%%
% Experimental section

\section{Experimental Methods}
\subsection{Samples}\label{sec:samples}
\FS single crystals have been grown by the chemical transport reaction method. As starting material we used pre-synthesized poly-crystalline powder prepared by solid state reaction from the high-purity elements
Fe (99.99\,\%) and Sn (99.995\,\%). Iodine was used as the transport agent.
The growth was performed in SiO$_2$ ampoules in two-zone furnaces at temperatures between 730 and 680\,K.
After 4-6 weeks of transport, the single-crystals, having a plate-like form  with a thickness in the range 20-40\,$\mu$m along the $c$-axis and 3-5\,mm in the $ab$-plane [see inset of Fig.\ref{fig1}(b)], were found in the hot part of ampoule.

The sample used for the experiments has a Residual Resistance Ratio of $RRR=29.3$ [Fig.\ref{fig1}(b)] indicating high crystalline quality. The derivative of the resistivity, $d\rho(T)/dT$ (green graph), indicates a qualitative change of the slope at approximately 100\,K where all spins are expected to be finally parallel to the $ab$-plane.
Fig.~\ref{fig1}(c) shows the temperature dependences of the in-plane and out-of-plane magnetic susceptibilities, $\chi_{ab}$ and $\chi_{c}$, respectively. For the analysis of the out-of-plane susceptibility, demagnetization effects had to be taken into account because of the thin-slab geometry of the sample. At high temperatures, $\chi_{c}$ is more than three times larger than $\chi_{ab}$. Toward 100\,K this difference diminishes. Below 100\,K $\chi_{c}$ starts to decrease while $\chi_{ab}$ saturates. Consequently, at low temperatures the $c$-axis becomes harder than the in-plane direction with $\chi_{ab}/\chi_{c}\approx 1.2$ at 2\,K.  In the same temperature range, where $\chi_{ab}$ and $\chi_{c}$ cross each other, the resistivity shows a kink, as best seen in its temperature derivative in Fig.~\ref{fig1}(b). After witnessing the spin reorientation in magnetic and transport properties, we turn to the Raman spectroscopic study of \FS, the main subject of this work.

\subsection{Light scattering}

\red{We performed polarized inelastic light scattering experiments at an excitation wavelength of 575\,nm (Coherent GENESIS MX-SLM577-500). The samples were attached to the cold finger of a $^4$He flow cryostat. Polarized photons hit the sample at an angle of incidence of $66^{\circ}$  yielding a spot size of approximately $50\times100\,\mu{\rm m}^2$. The polarized scattered photons were collected along the surface normal of the sample and focused on the entrance slit of a double monochromator. The resolution of the spectrometer is set at 2.8\,cm$^{-1}$. Polarized photons having the selected energy were recorded with a CCD detector. The number of photons per second is proportional to the Van-Hove function $S(q \approx 0, \Omega) = \hbar/\pi \{1+n(\Omega,T)\}\chi^{\prime\prime}(\Omega,T)$ where $n(\Omega,T)$ is the Bose factor and $\chi^{\prime\prime}$ is the imaginary part of Raman response function.}
 
For the measurements shown here we used only two polarization configurations, $RR$ and $RL$, where $R=2^{-1/2}(x+iy)$ and $L=2^{-1/2}(x-iy)$ for the incoming light (first symbol) inside the sample. \red{An absorbed laser power of $P_{\rm{abs}} = 4.0$\,mW independent of the polarization is used. $P_{\rm{abs}}$ induced a heating in the spot region of approximately 1-2\,K/mW.}  Here only the holder temperature is indicated.  Since we use (near) backscattering  configurations both signs change for the scattered light (second symbol). The details of the Raman selection rules will be discussed below. Due to the symmetric shape of the observed phonon modes and the narrow line width (FWHM), $\Gamma_{\rm L}(T) \ll \omega_0(T)$, where $\omega_0$ is the resonance energy of the respective mode, their line shapes may be described by temperature-dependent Lorentz functions.

\subsection{Selection rules and simulations}\label{sec:computational-methods}
The Raman-active phonon energies and eigenvectors at the $\Gamma$ point of the Brillouin zone were derived on the basis of the crystal structure using density functional theory (DFT). The symmetry selection rules may be determined from the point group and the atomic positions in the crystal. The space group of \FS is R$\bar{3}$m (No. $166$) and belongs to D$_{3d}$ $\red{(\bar{3}1m)}$ point group. The corresponding Raman tensors read

\begin{equation}\label{eq:A1g}
	A_{1g}=\left(
	\begin{array}{ccc}
		a  &  0  & 0\\
		0  &  a  & 0\\
		0  &  0  & b\\
	\end{array}
	\right),
\end{equation}

\begin{equation}\label{eq:Eg1}
	~~~~~~~~E_{g}^{(1)}=\left(
	\begin{array}{ccc}
		c  &  0  & 0\\
		0  &  -c  & d\\
		0  &  d  & 0\\
	\end{array}
	\right),~~\mathrm{and}
\end{equation}

\begin{equation}\label{eq:Eg2}
	~~~~E_{g}^{(2)}=\left(
	\begin{array}{ccc}
		0  &  -c  & -d\\
		-c  &  0 & 0\\
		-d  &  0  & 0\\
	\end{array}
	\right).
\end{equation}

According to the Wyckoff positions of the Fe (\red{$18h$}) and Sn atoms (\red{$6c$}) in \FS, there are 4 \Alg and 5 \Eg Raman-active phonons. On the basis of the Raman tensors [Eqs. (\ref{eq:A1g}), (\ref{eq:Eg1}), and (\ref{eq:Eg2})] the \Alg and  \Eg phonons, may be projected separately in the $RR$ and $RL$ channel, respectively.

Electronic structure calculations were carried out using DFT and the projector augmented wave (PAW) method as implemented in \red{VASP} ~\cite{kresse_ab_1994, kresse_efficiency_1996, kresse_efficient_1996, kresse_ultrasoft_1999}. The generalized gradient approximation as parameterized in the Perdew-Burke-Ernzerhof (PBE) functional~\cite{perdew_generalized_1996} was used to treat exchange and correlation effects. The cutoff for the plane-wave basis was chosen as $680$\,eV, and the Brillouin-zone was sampled with a $10 \times 10 \times 10$ $\Gamma$-centered Monkhorst-Pack grid. The $\mathrm{Fe_3Sn_2}$ crystal structure reported in~\cite{fenner_non-collinearity_2009} was fully relaxed until the forces on all atoms were below $0.001\,{\rm eV\cdot\AA^{-1}}$. $\Gamma$-point phonon calculations were performed using Density Functional Perturbation theory as implemented in \red{VASP}. For the symmetry analysis of all phonon modes the \textsc{phonopy} package \cite{togo_first_2015} was employed, allowing for an unambiguous assignment to the experimental frequencies, as compiled in Table~\ref{tab:fitting parameters}. \red{The eigenvectors characterizing the atomic displacement coordinates are listed in Supplementary Materials A.}\\

\begin{table*}
	\caption{Phonon energies and widths (FWHM) of \FS at 4.2\,K. \FS has four fully symmetric and five \Eg phonon modes four of which were observed experimentally. In addition to the theoretical and experimental energies $\omega_0$ and Lorentzian widths $\Gamma_{L}(T)$ the table displays also the phonon-phonon coupling parameters $\lambda_{i,{\rm ph-ph}}$ as derived from the approximative harmonic fits to the temperature dependent line widths. The two values of $\lambda_{1,{\rm ph-ph}}$ for \Alg(1) correspond to temperatures below and above 100\,K as indicated in Fig. \ref{fig:phonons-Ta1g}(a) by dashed-dotted and dotted lines, respectively.}
	\setlength{\tabcolsep}{2mm}{
		\begin{tabular}{ c c c c c c c c c c c}
			\specialrule{0em}{2pt}{2pt}
			\hline
			\specialrule{0em}{2pt}{2pt}
			% after \\: \hline or \cline{col1-col2} \cline{col3-col4} ...
			\multicolumn{2}{c} {Phonon} & \Alg(1) & \Alg(2) & \Alg(3) & \Alg(4) & \Eg(1) & \Eg(2) & \Eg(3) & \Eg(4)& \Eg(5) \\
			\specialrule{0em}{2pt}{2pt}
			\hline
			%\hline
			\specialrule{0em}{1.5pt}{1.5pt}
			\multirow{2}{*}{Energy (cm$^{-1}$)}&Simulation & 83.9 & 140.1 & 232.1 & 241.8 & 92.7 & 138.2 & 147.1 & 196.7 & 278.3\\
			\specialrule{0em}{1.5pt}{1.5pt}
			&Experiment & 86.6 & 146.8 & 237.7 & 251.6 & 94.3 & 133.8 & 147.0 & 199.8 & - \\
			\specialrule{0em}{1.5pt}{1.5pt}
			\multicolumn{2}{c} {FWHM (cm$^{-1}$)} &4.5 & 4.1 & 4.7 & 4.9 & 3.6 & 3.1 & 3.8 & 4.1 & - \\
			\specialrule{0em}{1.5pt}{1.5pt}
			%\bottomrule
			\multicolumn{2}{c} {$\lambda_{i, \rm{ph-ph}}$} &0.20; 0.05&0.14&0.68&0.74&0.11&1.11&0.29&0.42 & - \\
			\specialrule{0em}{1.5pt}{1.5pt}
			%\bottomrule
			\hline
	\end{tabular}}
	\label{tab:fitting parameters}
\end{table*}

\section{Results and Discussions}
The main focus of the paper is placed on the analysis of the phonon modes in the temperature range of the re-orientation of the Fe spins. Briefly, we will also discuss the electronic continuum.

\subsection{Phonons}
\begin{figure*}[ht]
	\centering
	\includegraphics[width=0.99\linewidth]{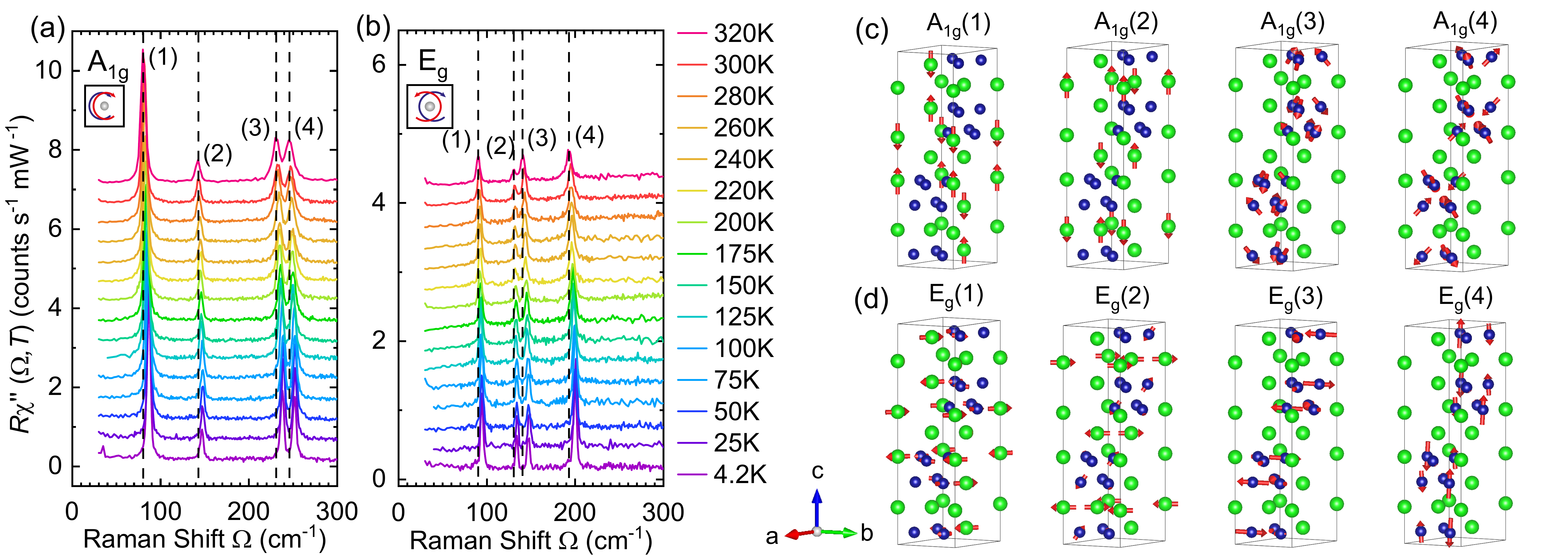}
	\caption{Raman spectra and phonon eigenvectors of \FS. (a) and (b) Temperature dependence of the eight observed phonon modes in \FS. The spectra at the lowest temperature have the experimental intensity. The \Alg spectra are consecutively offset by 0.5 cts/(s mW) and those in \Eg symmetry by 0.3 cts/(s mW) each for clarity. \red{The clockwise and counter-clockwise half circle arrows indicate the $R$ and $L$ polarization in (a) and (b), respectively.} Columns (a) and (b) separately display the \Alg- and \Eg phonon modes, respectively, (c) and (d) show the respective eigenvectors (small amplitudes omittted).}
	\label{fig:phonons-EV}
\end{figure*}

Figures~\ref{fig:phonons-EV}(a) and (b) show, respectively, the \Alg and \Eg Raman spectra at temperatures ranging from 4.2\,K to 320\,K. We can identify four \Alg phonons and four out of the expected five \Eg phonons. These phonons harden continuously with decreasing temperature. The absent fifth \Eg phonon might be too weak in intensity to be detected.

For a quantitative analysis the phonon peaks were fitted using a Voigt function, which is a convolution of the Lorentzian shape of the phonons and a Gaussian for the spectral resolution of the setup. In the narrow spectral range displayed in Fig.~\ref{fig:phonons-EV} and the laser line used, the resolution is constant and was set at 2.8\,\wn. Thus the widths indicated, directly correspond to those of the phonons. The reasonable quality of the fits (\red{see Fig. S1}) indicates that the Lorentzian widths $\Gamma_{L}(T)$ result from the finite life time of the phonons, determined by phonon-phonon decay into two modes having the same energy $\omega_0/2$ and opposite wave vectors $\vk$ and $-\vk$ \cite{Klemens:1966}.

\begin{figure*}[ht]
  \centering
  \includegraphics[width=0.9\linewidth]{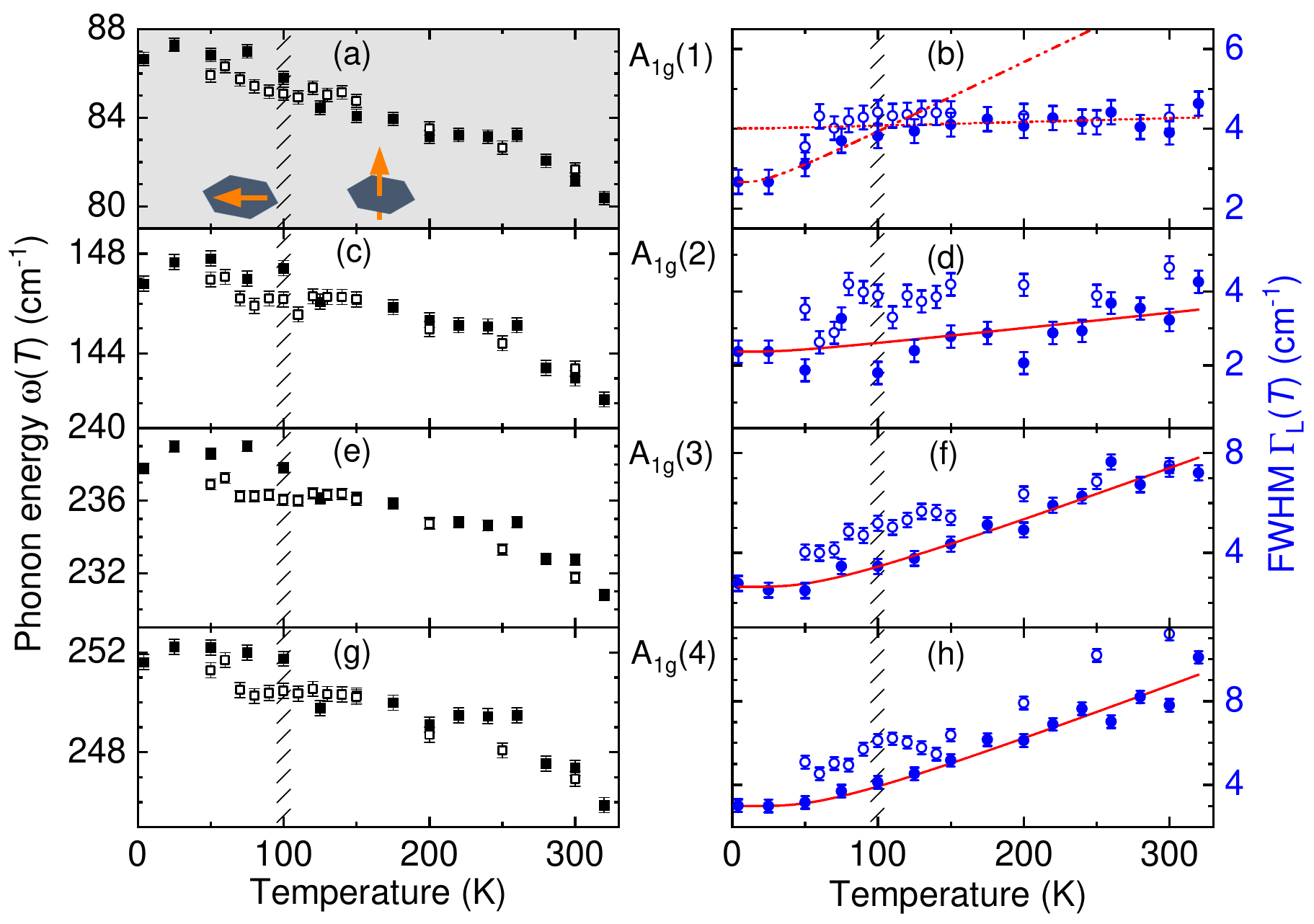}
  \caption{Phonon energies $\omega_0$ (black) and Lorentzian widths $\Gamma_L$ (blue) of \FS \red{in \Alg symmetry}. The average magnetization with respect to the $ab$-plane is indicated in (a). The parameters are derived by fitting procedures as described in the text. The first series of measurements (full black and blue symbols) was taken at temperatures \red{listed in the following sequence: [320, 300, 200, 100, 75, 50, 25, 4.2, 125, 150, 175, 220, 240, 260, 280]\,K}. The second series (open black and blue symbols) was measured in a row warming up from 50\,K with a distance of 10\,K in the interesting range. All linewidths $\Gamma_L$ were fitted according to equation (\ref{eqn:linewidth})  (red lines). The linewidth of the \Alg-mode at $86\,$cm$^{-1}$ in (a) shows anomalous behavior, displaying a change in the expected temperature dependence at approximately $100\,$K as indicated by the hatched area. Eq. (\ref{eqn:linewidth}) yields $\lambda_{1,{\rm ph-ph}}=0.2$ at low temperature and only 0.05 above 100\,K. }
	\label{fig:phonons-Ta1g}
\end{figure*}

\begin{figure*}[ht]
  \centering
  \includegraphics[width=0.9\linewidth]{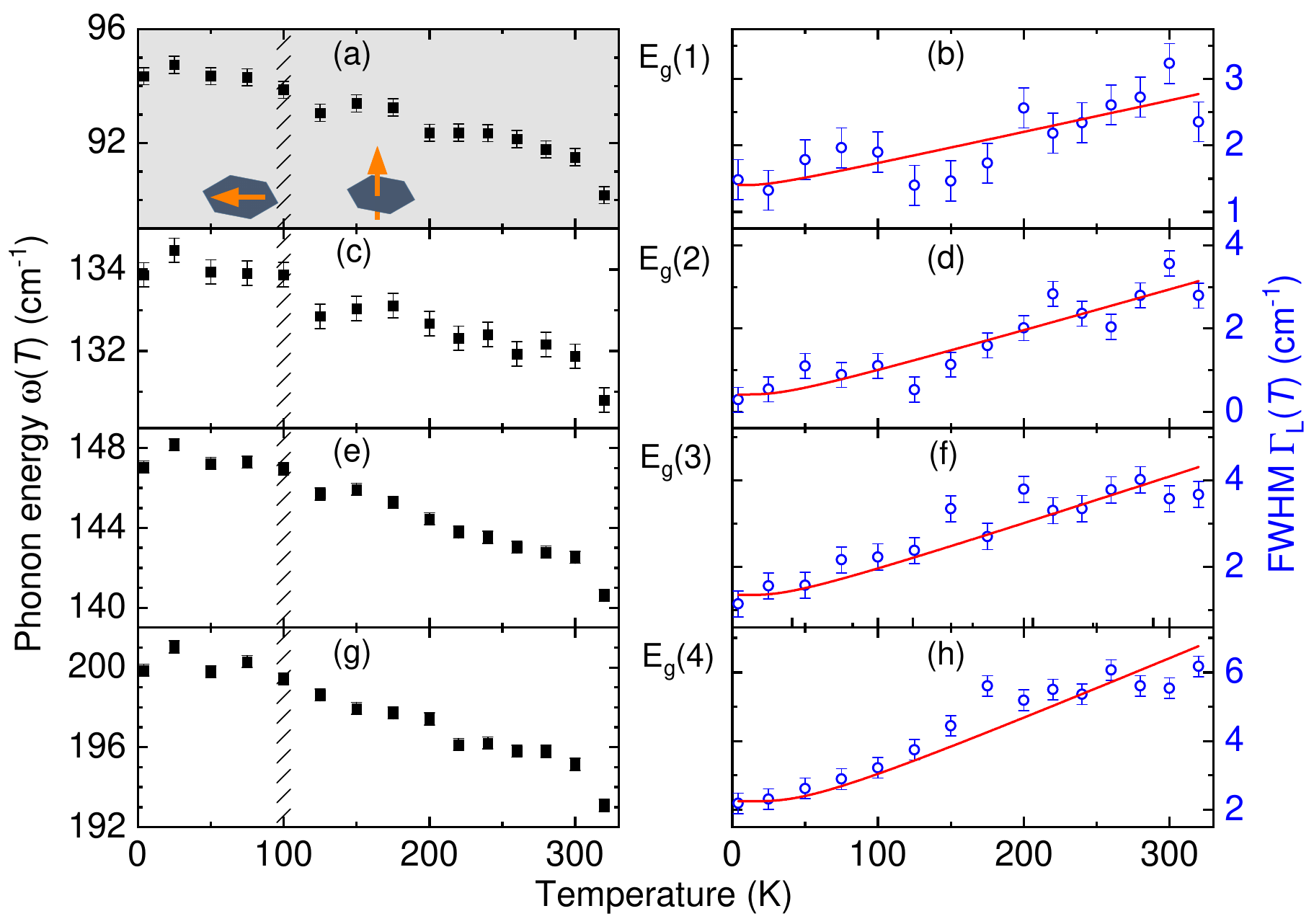}
  \caption{\red{Phonon energies $\omega_0$ (black) and Lorentzian widths $\Gamma_L$ (blue) of \FS in \Eg symmetry. The linewidths and phonon energies behave like usual phonons without anomaly near the spin reorientation transition temperature.}}
	\label{fig:phonons-Teg}
\end{figure*}

The peak energies and linewidths (FWHM) derived in this way are depicted in Fig. \ref{fig:phonons-Ta1g} and \ref{fig:phonons-Teg}, and labeled consecutively from low to high. Their respective values at $4.2$\,K are listed in Table \ref{tab:fitting parameters} and found to be in good agreement with the simulation. All phonon modes become harder and narrower upon cooling. The usual changes in width and energy are related to the unharmonic decay \cite{Klemens:1966} and the lattice contraction \cite{Postmus:1968}. However, as opposed to the \Eg modes some of the \Alg modes show weak but significant deviations from the expected behavior in the temperature range around 100\,K. In particular, kinks are observed in the energies of modes 1, 3, and 4 [see hatched range in Fig. \ref{fig:phonons-Ta1g}(a) and (c)]. After remeasuring the \Alg spectra [\red{open} symbols in Fig. \ref{fig:phonons-Ta1g}(a)--(d)] the kinks can still be observed but are shifted slightly. The origin of this shift is not entirely clear but may be related to the first-order nature of the reorientation transition and the then expected hysteresis \cite{Kumar:2019}. In fact, the \red{open} points were obtained upon heating from low to high temperature in an uninterrupted series whereas the black points were measured \red{upon cooling as indicated in the figure caption}. More experiments are required to finally clarify this issue.

The conventional reduction of the phonon widths with decreasing temperature can be understood in terms of the anharmonic decay described above \cite{Klemens:1966} and may be represented by

\begin{equation}
	\Gamma_L(T) = \Gamma_{L,0} \Biggl(1+\frac{2\lambda_{\rm ph-ph}}{\exp(\frac{\hbar\omega_0}{2k_BT})-1}\Biggr).
	\label{eqn:linewidth}
\end{equation}

$\Gamma_{L,0}(T)$ and $\omega_0$ in Eq. (\ref{eqn:linewidth}) can be extrapolated to zero from the experimental points below $50\,$K, leaving only the phonon-phonon coupling $\lambda_{i,{\rm ph-ph}}$ as a free parameter. The corresponding curves are indicated in red in Fig. \ref{fig:phonons-Ta1g} and \ref{fig:phonons-Teg}, and the resulting values for $\lambda_{\rm ph-ph}$ are compiled in Table \ref{tab:fitting parameters}. \red{As opposed to the linewidth the phonon energy depends on the occupation and the thermal expansion where the latter one is approximately larger by two orders of magnitude and depends crucially on the variation of the interatomic potential with distance \cite{Postmus:1968}. The temperature dependences of the phonon width and energy are related by the usual Kramers-Kronig transformation and were determined for constant volume by Balkanski and coworkers \cite{Balkanski:1983}. However, in an experiment the pressure rather than the volume are constant, and this approach is not applicable (see, e.g., Ref. \cite{Eiter:2014a}).Thus,  the analysis of the phonon energy requires the knowledge of the thermal expansion and the Gr\"uneisen parameters which are currently not available.} \\
All phonons except for one show essentially conventional behavior and become exponentially narrower upon cooling. Only the \Alg(1) phonon at 86\,\wn shows unexpected behavior directly in the data, and the description of the temperature dependence according to Eq. (\ref{eqn:linewidth}) is similar to the other modes only for T $\leq 100\,$K [dashed-dotted in Fig. \ref{fig:phonons-Ta1g}(a)] and yields $\lambda_{1,{\rm ph-ph}}=0.2$. For T $\geq 100\,$K (dotted) the width is almost temperature independent corresponding to an unusually small coupling of $\lambda_{1,{\rm ph-ph}}=0.05$. So we conclude that the symmetric phonon-phonon decay channel is essentially blocked when the spins point along the $c$-axis and becomes accessible only when the spins are rotated into the $ab$-plane. This hand-waving argument certainly needs further theoretical analysis but a realistic model of the spin-phonon coupling is beyond the scope of this paper and requires complex, presumably numerical work.

%It should be pointed out that the phonon energy changes are mainly contributed from phonon-phonon scattering and thermal expansion. In general, the latter one has approximately two orders of magnitude larger contribution than the anharmonic decay \cite{Postmus:1968, Eiter:2014a}. Since we do not have the thermal expansion results, the fit of the T-dependent phonon energy is impossible.
\subsection{\Eg continuum at low energy}
Since \FS is purely metallic and orders at \TC$\approx670$\,K, gaps or isolated electronic or magnetic modes are not immediately expected in the temperature range studied. Due to the magnetic anisotropy one may expect to see a single magnon at finite energy given by the anisotropy field. However, our microwave experiments show that the energy is presumably too small \red{for the Raman experiment}. In addition, fluctuations as a consequence  of geometric frustration may smear out the energies.

\begin{figure}[ht]
  \centering
  \includegraphics[width=0.5\linewidth]{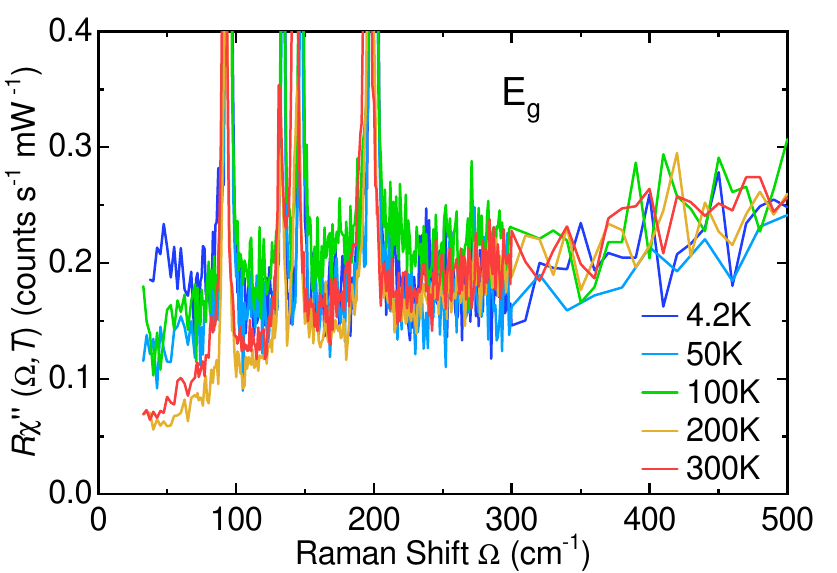}
  \caption{Electronic continuum in \Eg symmetry at temperatures as indicated. \red{The factor $R$ is a constant of proportionality which absorbs all experimental factors. The contribution from the laser line is negligible above 30\,\wn. The intensity below 150\,\wn increases continuously upon cooling which implies a concomitant increase of the initial slope indicating a metallic ground state in agreement with the transport result.} The increase at the limit $\Omega\to0$ is a real effect since the spectra at 4.2\,K have a peak at finite energy and then decrease again as opposed to a laser-induced divergence.}
	\label{fig:continuum}
\end{figure}

The small variations with temperature of the \Eg spectra below 100\,cm$^{-1}$, as shown in Fig. \ref{fig:continuum}, look unspectacular at first glance. However, the increase at low temperature must be considered real since a peak at approximately 47\,\wn can be resolved at 4.2\,K rather than a divergence towards zero energy as in the case of diffuse scattering of the laser light. The \red{intensity} increase may either originate from particle-hole excitations reflecting the temperature dependence of the resistivity [see Fig. \ref{fig1}(b)] or from interband transitions as suggested by the optical conductivity \cite{Biswas:2020} or from fluctuations similar to FeSe \cite{Baum:2019}. \red{For clarification larger crystals with flat surfaces are necessary.}

\red{Compared with the \Eg spectra, the \Alg continuum is essentially temperature-independent (see Supplementary Materials C). As in the case of the \Eg spectra, the data are reliable above 30\,\wn. Obviously, different excitations or regions of the Brillouin zone are projected in the two symmetries. Given the correspondence between transport and low-energy \Eg spectra, the most likely explanation is that the carrier relaxation observed in the \Alg spectra is almost temperature independent. This interesting anisotropy calls for more studies in optimized samples.} 

\section{Conclusions}
We have studied the Kagome ferromagnet \FS using polarized Raman scattering. We identified all but one phonon lines and compared them to simulations of the lattice dynamics on the basis of density functional theory. The agreement between predicted and experimental phonon energies is better than 5\,\% for all lines. All phonons broaden upon heating. By assuming symmetric decay of the lines at $\omega_0$ and $|\vq|\approx 0$ into two lines at $\omega_0/2$ and $\pm \vk$ the dimensionless phonon-phonon coupling $\lambda_{i,{\rm ph-ph}}$ was extracted and found to be in the range 0.1 to 1.1 for all lines and temperatures with one exception: The \Alg phonon with the lowest energy exhibits $\lambda_{1,{\rm ph-ph}}=0.2$ below 100\,K and $\lambda_{1,{\rm ph-ph}}=0.05$ above 100\,K thus it has only very small phonon-phonon coupling at temperatures with the spins aligned along the $c$-axis parallel to the motion of the Sn atoms for the \Alg(1) mode. For this specific eigenvector the coupling to the Fe spins parallel to the $c$-axis is certainly small, and one may speculate that the relatively large $\lambda_{i,{\rm ph-ph}}$ values of the other phonons result from coupling \textit{via} the spins. As soon as the spins rotate into the plane also the \Alg(1) mode couples to the spins and thus participates in the anharmonic decay.

Because of the relatively high Curie temperature of \TC$=670$\,K in \FS one does not expect strong changes in the electronic properties below room temperature. Yet, the \Eg continuum exhibits a substantial variation with temperature at low energies which are reminiscent of the strong temperature dependence of the resistivity $\rho(T)$. However, the low-energy peak in the \Eg spectra may also originate from a band gap induced by the spin reorientation as suggested by optical measurements \cite{Biswas:2020}, although the rather different energies in the two experiments argues against this possibility, or from magnetic fluctuations in a geometrically frustrated system. There are no indications of a flat band in the 200\,meV range.  Yet, with the available equipment we cannot obtain a sufficiently high data quality for a quantitative analysis.

%\begin{figure*}[ht]
%  \centering
%  \includegraphics[width=0.9\linewidth]{figA1_He}
%  \caption{Experimental phonons (points) along with the Voigt fits (full red lines) as described in the text. The Gaussian width was set at  2.8\,\wn.}
%	\label{fig:Voigt}
%\end{figure*}
\medskip

% Acknowledgements
\medskip
\textbf{Acknowledgements} \par %delete if not applicable))
We thank Y.-F. Xu for fruitful discussions. This work is supported by the Deutsche Forschungsgemeinschaft (DFG) through the coordinated programme TRR80 (Project-ID 107745057) and project HA 2071/12-1. G. He would like to thank the Alexander von Humboldt Foundation for support through a fellowship.

\textbf{Author contributions} \par
G. He and L. Peis contributed equally to the Raman study. 
% References
\medskip

% Use the following code if you wish to generate your bibliography with BibTeX;
% replace the string "MSP-template" below with the name(s) of
% the BibTeX data base(s) you want to use.
% The resulting bibliography-output (the content of the .bbl file)
% must be pasted back into this file before submission.
% Please also include your BibTeX data base file(s) in your submission
% so that we can re-run BibTeX if necessary.
%
\bibliographystyle{MSP}
\bibliography{literatureR3}

\begin{thebibliography}{10}
\providecommand{\url}[1]{\texttt{#1}}
\providecommand{\urlprefix}{URL }

\bibitem{Trumpy:1970}
G.~Trumpy, E.~Both, C.~Dj\'ega-Mariadassou, P.~Lecocq,
\newblock \emph{Phys. Rev. B} \textbf{1970}, \emph{2} 3477.

\bibitem{LeCaer:1978}
G.~Le~Ca\"er, B.~Malaman, B.~Roques,
\newblock \emph{J. Phys. F: Met. Phys.} \textbf{1978}, \emph{8} 323.

\bibitem{Malaman:1978}
B.~Malaman, C.~Fruchart, G.~Le~Ca\"er,
\newblock \emph{J. Phys. F: Met. Phys.} \textbf{1978}, \emph{8} 2389.

\bibitem{fenner_non-collinearity_2009}
L.~A. Fenner, A.~A. Dee, A.~S. Wills,
\newblock \emph{J. Phys. Condens. Matter} \textbf{2009}, \emph{21} 452202.

\bibitem{HouZP:2017}
Z.~Hou, W.~Ren, B.~Ding, G.~Xu, Y.~Wang, B.~Yang, Q.~Zhang, Y.~Zhang, E.~Liu,
  F.~Xu, W.~Wang, G.~Wu, X.~Zhang, B.~Shen, Z.~Zhang,
\newblock \emph{Adv. Mater.} \textbf{2017}, \emph{29} 1701144.

\bibitem{Hou:2018}
Z.~Hou, Q.~Zhang, G.~Xu, C.~Gong, B.~Ding, Y.~Wang, H.~Li, E.~Liu, F.~Xu,
  H.~Zhang, Y.~Yao, G.~Wu, X.-x. Zhang, W.~Wang,
\newblock \emph{Nano Lett.} \textbf{2018}, \emph{18} 1274.

\bibitem{Tang:2020a}
J.~Tang, Y.~Wu, L.~Kong, W.~Wang, Y.~Chen, Y.~Wang, Y.~Soh, Y.~Xiong, M.~Tian,
  H.~Du,
\newblock \emph{Natl. Sci. Rev.} \textbf{2020}, \emph{8} nwaa200.

\bibitem{Tang:2020b}
J.~Tang, L.~Kong, Y.~Wu, W.~Wang, Y.~Chen, Y.~Wang, J.~Li, Y.~Soh, Y.~Xiong,
  M.~Tian, H.~Du,
\newblock \emph{ACS Nano} \textbf{2020}, \emph{14} 10986.

\bibitem{YinJX:2018}
J.-X. Yin, S.~S. Zhang, H.~Li, K.~Jiang, G.~Chang, B.~Zhang, B.~Lian, C.~Xiang,
  I.~Belopolski, H.~Zheng, T.~A. Cochran, S.-Y. Xu, G.~Bian, K.~Liu, T.-R.
  Chang, H.~Lin, Z.-Y. Lu, Z.~Wang, S.~Jia, W.~Wang, M.~Z. Hasan,
\newblock \emph{Nature} \textbf{2018}, \emph{562} 91.

\bibitem{WangQ:2016}
Q.~Wang, S.~Sun, X.~Zhang, F.~Pang, H.~Lei,
\newblock \emph{Phys. Rev. B} \textbf{2016}, \emph{94} 075135.

\bibitem{Biswas:2020}
A.~Biswas, O.~Iakutkina, Q.~Wang, H.~C. Lei, M.~Dressel, E.~Uykur,
\newblock \emph{Phys. Rev. Lett.} \textbf{2020}, \emph{125} 076403.

\bibitem{Lin:2020}
Z.-Z. Lin, X.~Chen,
\newblock \emph{phys. stat. solidi (RRL) Rapid Research Letters} \textbf{2020},
  \emph{14} 1900705.

\bibitem{Lin:2018}
Z.~Lin, J.-H. Choi, Q.~Zhang, W.~Qin, S.~Yi, P.~Wang, L.~Li, Y.~Wang, H.~Zhang,
  Z.~Sun, L.~Wei, S.~Zhang, T.~Guo, Q.~Lu, J.-H. Cho, C.~Zeng, Z.~Zhang,
\newblock \emph{Phys. Rev. Lett.} \textbf{2018}, \emph{121} 096401.

\bibitem{Nagaosa:2010}
N.~Nagaosa, J.~Sinova, S.~Onoda, A.~H. MacDonald, N.~P. Ong,
\newblock \emph{Rev. Mod. Phys.} \textbf{2010}, \emph{82} 1539.

\bibitem{Franz:2014d}
C.~Franz, F.~Freimuth, A.~Bauer, R.~Ritz, C.~Schnarr, C.~Duvinage, T.~Adams,
  S.~Bl\"ugel, A.~Rosch, Y.~Mokrousov, C.~Pfleiderer,
\newblock \emph{Phys. Rev. Lett.} \textbf{2014}, \emph{112} 186601.

\bibitem{Mielke:1991b}
A.~Mielke,
\newblock \emph{J. Phys. A} \textbf{1991}, \emph{24} 3311.

\bibitem{Mielke:1991a}
A.~Mielke,
\newblock \emph{J. Phys. A} \textbf{1991}, \emph{24} L73.

\bibitem{Mielke:1992}
A.~Mielke,
\newblock \emph{J. Phys. A} \textbf{1992}, \emph{25} 4335.

\bibitem{Nagaoka:1966}
Y.~Nagaoka,
\newblock \emph{Phys. Rev.} \textbf{1966}, \emph{147} 392.

\bibitem{Pollmann:2008}
F.~Pollmann, P.~Fulde, K.~Shtengel,
\newblock \emph{Phys. Rev. Lett.} \textbf{2008}, \emph{100} 136404.

\bibitem{Yin:2011}
Z.~P. Yin, K.~Haule, G.~Kotliar,
\newblock \emph{Nature Mater.} \textbf{2011}, \emph{10} 932.

\bibitem{Leonov:2015}
I.~Leonov, S.~L. Skornyakov, V.~I. Anisimov, D.~Vollhardt,
\newblock \emph{Phys. Rev. Lett.} \textbf{2015}, \emph{115} 106402.

\bibitem{Stadler:2015}
K.~M. Stadler, Z.~P. Yin, J.~von Delft, G.~Kotliar, A.~Weichselbaum,
\newblock \emph{Phys. Rev. Lett.} \textbf{2015}, \emph{115} 136401.

\bibitem{Baum:2019}
A.~Baum, H.~N. Ruiz, N.~Lazarevi\'c, Y.~Wang, T.~B\"ohm,
  R.~Hosseinian~Ahangharnejhad, P.~Adelmann, T.~Wolf, Z.~V. Popovi\'c,
  B.~Moritz, T.~P. Devereaux, R.~Hackl,
\newblock \emph{Commun. Phys.} \textbf{2019}, \emph{2} 14.

\bibitem{Eiter:2014a}
H.-M. Eiter, P.~Jaschke, R.~Hackl, A.~Bauer, M.~Gangl, C.~Pfleiderer,
\newblock \emph{Phys. Rev. B} \textbf{2014}, \emph{90} 024411.

\bibitem{kresse_ab_1994}
G.~Kresse, J.~Hafner,
\newblock \emph{Phys. Rev. B} \textbf{1994}, \emph{49} 14251.

\bibitem{kresse_efficiency_1996}
G.~Kresse, J.~Furthmueller,
\newblock \emph{Comput. Mater. Sci.} \textbf{1996}, \emph{6} 15.

\bibitem{kresse_efficient_1996}
G.~Kresse, J.~Furthmueller,
\newblock \emph{Phys. Rev. B} \textbf{1996}, \emph{54} 11169.

\bibitem{kresse_ultrasoft_1999}
G.~Kresse, D.~Joubert,
\newblock \emph{Phys. Rev. B} \textbf{1999}, \emph{59} 1758.

\bibitem{perdew_generalized_1996}
J.~P. Perdew, K.~Burke, M.~Ernzerhof,
\newblock \emph{Phys. Rev. Lett.} \textbf{1996}, \emph{77} 3865.

\bibitem{togo_first_2015}
A.~Togo, I.~Tanaka,
\newblock \emph{Scr. Mater.} \textbf{2015}, \emph{108} 1.

\bibitem{Klemens:1966}
P.~G. Klemens,
\newblock \emph{Phys. Rev.} \textbf{1966}, \emph{148} 845.

\bibitem{Postmus:1968}
C.~Postmus, J.~R. Ferraro, S.~S. Mitra,
\newblock \emph{Phys. Rev.} \textbf{1968}, \emph{174} 983.

\bibitem{Kumar:2019}
N.~Kumar, Y.~Soh, Y.~Wang, Y.~Xiong,
\newblock \emph{Phys. Rev. B} \textbf{2019}, \emph{100} 214420.

\bibitem{Balkanski:1983}
M.~Balkanski, R.~F. Wallis, E.~Haro,
\newblock \emph{Phys. Rev. B} \textbf{1983}, \emph{28} 1928.

\end{thebibliography}

\end{document}